**TITLE:**

Joint k-TE Space Image Reconstruction and Data Fitting for T2 Mapping

**RUNNING TITLE:**

Joint k-TE Reconstructed T2 Mapping

**AUTHORS:**

Yan Dai B.S. [1], Xun Jia Ph.D.[2], Yen-Peng Liao, PhD[1], Jiaen Liu, PhD[3], Jie Deng Ph.D.[1]

[1] Department of Radiation Oncology, University of Texas Southwestern Medical Centre, TX, USA

[2] Department of Radiation Oncology and Molecular Radiation Sciences, Johns Hopkins University, MD, USA

[3] Advanced Imaging Research Center, University of Texas Southwestern Medical Centre, TX, USA

**CORRESPONDENCE:**

Jie Deng, Ph.D.

214-645-5140

Department of Radiation Oncology,

University of Texas Southwestern Medical Centre, 2280 Inwood Rd, Dallas.

Email: Jie.Deng@UTSouthwestern.edu

**ACKNOWLEDGEMENT**

This work was supported in part by the Cancer Prevention and Research Institute of Texas (grant \#RP200573) and the National Cancer Institute (grant \#R01CA227289).

## ABSTRACT

**Objectives:** To develop a joint k-TE reconstruction algorithm to reconstruct the T2-weighted (T2W) images and T2 map simultaneously.

**Materials and Methods:** The joint k-TE reconstruction model was formulated as an optimization problem subject to a self-consistency condition of the exponential decay relationship between the T2W images and T2 map. The objective function included a data fidelity term enforcing the agreement between the solution and the measured k-space data, together with a spatial regularization term on image properties of the T2W images. The optimization problem was solved using Alternating-Direction Method of Multipliers (ADMM). We tested the joint k-TE method in phantom data and healthy volunteer scans with fully-sampled and under-sampled k-space lines. Image quality of the reconstructed T2W images and T2 map, and the accuracy of T2 measurements derived by the joint k- TE and the conventional signal fitting method were compared.

**Results:** The proposed method improved image quality with reduced noise and less artifacts on both T2W images and T2 map, and increased measurement consistency in T2 relaxation time measurements compared with the conventional method in all data sets.

**Conclusions:** The proposed reconstruction method outperformed the conventional magnitude image-based signal fitting method in image quality and stability of quantitative T2 measurements.

## 1. INTRODUCTION

T2-weighted (T2W) MR images provide an important image contrast mechanism resulting from the difference of tissue transverse relaxation time for anatomical delineation, disease diagnosis, and tissue characterization[1,2]. The T2 relaxation time measurement derived from a series of T2W images acquired at different echo-times (TEs) based on a mono-exponential signal decay model provided a quantitative method for tissue characterization, which has been used for cancer diagnosis and treatment response assessment [3-5]. Pixel-by-pixel T2 measurements, called T2 mapping, has been integrated into various clinical MRI protocols such as cardiovascular MRI for the diagnosis of myocardial inflammation and edema[6-9], and neuroimaging for brain maturation evaluation [10-14]. The conventional T2 mapping reconstruction method includes a two-step process, in which the magnitude T2W image acquired at each TE is reconstructed from its own k-space data, e.g., by Fast Fourier Transform (FFT), followed by a pixel-by-pixel fitting of the T2W MR signal decay to derive the T2 map. The accuracy of T2 measurement using the conventional 2-step method is limited by low signal-to-noise-ratio (SNR), low spatial resolution, motion artifacts, as well as image quality degradation[5,15,16]. Acquiring a full set of T2W image based on spin echo (SE) or fast spin echo (FSE) pulse sequences takes a relatively long imaging time, resulting in pixel misregistration and thus inaccurate signal fitting. Other types of MRI pulse sequences have also been used for T2 mapping such as T2-prepared balanced steady-state free precession (bSSFP), and Gradient And Spin Echo (GraSE)[17], a hybrid technique that acquires a series of gradient echoes and spin echoes from a train of 180° radiofrequency pulses, were used for fast myocardial T2 mapping. In addition, various types of post-processing algorithms for deriving T2 map based on reconstructed magnitude T2W images have been used but with the problems of noisy data fitting and reproducibility[155].

Parallel imaging techniques were introduced to accelerate MRI acquisition by estimating missing data through coil-based calibration. The generalized auto-calibrating partially parallel acquisitions(GRAPPA)[18] is one of the parallel imaging methods that uses the linear relationship between the acquired k-space lines and adjacent missing lines as a constraint in the block-wise calibrations across all coil elements to estimate the missing data. This constraint was further developed to calibrate between every single k-space data point and its neighboring data points across all coil elements using a matrix called SPiRIT (iterative self-consistent parallel imaging reconstruction) operator[19]. Similarly, spatial sparsity of an MR image was considered as prior knowledge in image domain to reconstruct under-sampled data, namely compressed sensing (CS). The most common form of CS-based reconstruction is to minimize a loss function consisting of a data fidelity term and a specifically designed regularization term that penalizes violations the sparsity assumption[20-22]. Essentially, MRI reconstruction can be generally viewed as an inverse problem that aims to restore data from non-ideal measurements. Prior knowledge is useful in data correction and constraining solutions as used in the above-mentioned parallel imaging and CS techniques. More recently, neural network demonstrated good performance in MR image reconstruction[23-25]. A U-net was proposed to reconstruct under-sampled T2W images given the corresponding T1-weighted (T1W) images[26]. With the flexibility of neural network, the sparsity between T2W images at different TEs and that between T2W image and T1W image can be well integrated into the reconstruction of T2W images[27].

In this study, we propose a joint k-TE space reconstruction method that exploits structural correlations between different T2W images acquired at different TEs and the mono-exponential decay relationship between T2W images and the corresponding T2 map as the regularization terms[28]. Furthermore, the joint k-TE reconstruction algorithm can be applied to under-sampled

T2W images, in which CS was not only used in each T2W image itself but also between different T2W images[27,29-31]. This work aims to reconstruct T2W images and T2 mapping simultaneously by solving an optimization problem. The error in an T2W image was iteratively corrected by the information obtained from other T2W images at different TEs as well as from the corresponding T2 map by enforcing the exponential decay constraint, which in turn also reduced the noise in T2 map. In addition, an image denoising filter algorithm was applied to T2W images as a prior to restore image smoothness via the plug-and-play approach. We tested this joint k-TE reconstruction method in both phantom data and healthy volunteer images, and compared the image quality and accuracy of T2 measurements with those obtained by the conventional method for fully sampled data and CS-based method for under-sampled data.

## 2. MATERIALS AND METHODS

### 2.A. Model for joint reconstruction and data fitting

Let us denote the measured k-space data at $TE_i$ as $\boldsymbol{g}(z, TE_i)$ and the magnitude image to be reconstructed as $\boldsymbol{f}(x, TE_i)$, with $x$ being the spatial coordinate. There exists a Fourier transform relationship between them:

$$\boldsymbol{g}(z, TE_i) = \boldsymbol{SFA}(x, TE_i)\boldsymbol{f}(x, TE_i) + \boldsymbol{n} \qquad \textbf{eq. 1}$$

where $\boldsymbol{n}$ denotes noise signal in data acquisition. $\boldsymbol{S}$ is the sampling matrix**,** $\boldsymbol{F}$ is the Fourier transform operator. $\boldsymbol{A}(x, TE_i)$ is the phase image. In this study, we assumed this is known and it was estimated from the image obtained by inverse Fourier transform of $\boldsymbol{g}(z, TE_i)$ by taking its phase factors. The T2W MR magnitude images at different TEs are related by the T2 relaxation model:

$$\boldsymbol{f}(x, TE_i) = \boldsymbol{f}(x, TE_0) e^{-\frac{TE_i - TE_0}{\boldsymbol{T2}(x)}} \quad \textbf{eq. 2}$$

where $\boldsymbol{T2}(x)$ is the T2 map.

We proposed to solve the following optimization model to jointly estimate the image $\boldsymbol{f}(x, TE)$ and the T2 map $\boldsymbol{T2}(x)$:

$$\{\boldsymbol{f}^*, \boldsymbol{T2}^*\} = \underset{\boldsymbol{f}, \mathbf{T2}}{\operatorname{argmin}} \sum_i |\boldsymbol{SFA}(x, TE_i)\boldsymbol{f}(x, TE_i) - \boldsymbol{g}(z, TE_i)|^2 + R[\boldsymbol{f}(x, TE_i), \lambda],$$

$$s.t.\ \boldsymbol{f}(x, TE_i) = \boldsymbol{f}(x, 0) e^{-\frac{TE_i - TE_0}{\boldsymbol{T2}(x)}}, \text{for } i > 0 \quad \textbf{eq. 3}$$

There are two terms in the objective function. The first one is a data fidelity term that ensures the agreement between the reconstructed magnitude images $\boldsymbol{f}(x, TE_i)$ and the corresponding measurements $\boldsymbol{g}(z, TE_i)$. The second term is employed to provide regulation on MR image in the spatial domain, where $\lambda$ is the parameter in the regularization function. In this study, we used the Block-matching and 3D filtering (BM3D) method for regularization[32], which was employed via the plug-and-play (PnP) approach presented in the next subsection. The constraint posts the connection among MR images $\boldsymbol{f}(x, TE_i)$ and the T2 map $\boldsymbol{T2}(x)$.

### 2.B. Numerical algorithm and implementation

The Alternating Direction Method of Multipliers was employed to solve this optimization problem[33]. As such, we first considered the optimization problem equivalent to **eq. *3*** by introducing a variable $\boldsymbol{v}$:

$$\{\boldsymbol{f}, \boldsymbol{T2}, \boldsymbol{v}\} = \underset{\boldsymbol{f}, \boldsymbol{T2}, \boldsymbol{v}}{\operatorname{argmin}} \frac{1}{2} \sum_i |\boldsymbol{SFA}(x, TE_i)\boldsymbol{f}(x, TE_i) - \boldsymbol{g}(z, TE_i)|^2 + R[\boldsymbol{v}(x, TE_i), \lambda],$$

**eq. 4**

$$s.t.\, \boldsymbol{f}(x, TE_i) = \boldsymbol{f}(x, TE_0) e^{-\frac{TE_i - TE_0}{\boldsymbol{T2}(x)}}, \text{for } i > 0\,;\ \boldsymbol{v}(x, TE_i) = \boldsymbol{f}(x, TE_i).$$

The augmented Lagrangian of this problem is

$$L_\rho = \frac{1}{2} \sum_i |\boldsymbol{SFA}(x, TE_i)\boldsymbol{f}(x, TE_i) - \boldsymbol{g}(z, TE_i)|^2 + R[\boldsymbol{v}(x, TE_i), \lambda]$$

$$+ \sum_{i>0} \boldsymbol{y}_i^T \left[ \boldsymbol{f}(x, TE_i) - \boldsymbol{f}(x, TE_0) e^{-\frac{TE_i - TE_0}{\boldsymbol{T2}(x)}} \right]$$

**eq. 5**

$$+ \sum_{i>0} \frac{\rho}{2} \left| \boldsymbol{f}(x, TE_i) - \boldsymbol{f}(x, TE_0) e^{-\frac{TE_i - TE_0}{\boldsymbol{T2}(x)}} \right|^2$$

$$+ \sum_i z_i^T [\boldsymbol{v}(x, TE_i) - \boldsymbol{f}(x, TE_i)] + \sum_i \frac{\rho}{2} |\boldsymbol{v}(x, TE_i) - \boldsymbol{f}(x, TE_i)|^2$$

where $\boldsymbol{y}_i$, $\boldsymbol{z}_i$ and $\rho$ are variables introduced in the algorithm. The ADMM solved the optimization problem by iteratively performing the following steps with $k$ being the index of iteration:

$$\boldsymbol{f}^{(k+1)}(x, TE_i) = \underset{\boldsymbol{f}(x, TE_i)}{\operatorname{argmin}} \frac{1}{2} |\boldsymbol{SFA}(x, TE_i)\boldsymbol{f}(x, TE_i) - \boldsymbol{g}(z, TE_i)|^2$$

$$+ \boldsymbol{y}_i^T \left[ \boldsymbol{f}(x, TE_i) - \boldsymbol{f}^{(k)}(x, TE_0) e^{-\frac{TE_i - TE_0}{\boldsymbol{T2}^{(k)}(x)}} \right]$$

$$+ \frac{\rho}{2} \left| \boldsymbol{f}(x, TE_i) - \boldsymbol{f}^{(k)}(x, TE_0) e^{-\frac{TE_i - TE_0}{\boldsymbol{T2}^{(k)}(x)}} \right|^2$$

**eq. 6**

$$+ \boldsymbol{z}_i^T \left[ \boldsymbol{v}^{(k)}(x, TE_i) - \boldsymbol{f}(x, TE_i) \right] + \frac{\rho}{2} \left| \boldsymbol{v}^{(k)}(x, TE_i) - \boldsymbol{f}(x, TE_i) \right|^2,$$

for $i > 0$

$$\boldsymbol{f}^{(k+1)}(x, TE_0) = \underset{\boldsymbol{f}(x,TE_0)}{\operatorname{argmin}} \frac{1}{2} |\boldsymbol{SFA}(x, TE_0)\boldsymbol{f}(x, TE_0) - \boldsymbol{g}(z, TE_0)|^2$$

$$+ \sum_{i>0} \boldsymbol{y}_{\boldsymbol{i}}^T \left[ \boldsymbol{f}^{(k+1)}(x, TE_i) - \boldsymbol{f}(x, TE_0) e^{-\frac{TE_i - TE_0}{\boldsymbol{T2}^{(k)}(x)}} \right]$$

$$+ \sum_{i>0} \frac{\rho}{2} \left| \boldsymbol{f}^{(k+1)}(x, TE_i) - \boldsymbol{f}(x, TE_0) e^{-\frac{TE_i - TE_0}{\boldsymbol{T2}^{(k)}(x)}} \right|^2 + \boldsymbol{z}_{\boldsymbol{i}}^T [\boldsymbol{v}^{(k)}(x, TE_0)$$

$$- \boldsymbol{f}(x, TE_0)] + \frac{\rho}{2} \left| \boldsymbol{v}^{(k)}(x, TE_0) - \boldsymbol{f}(x, TE_0) \right|^2$$

**eq. 7**

$$\frac{1}{\boldsymbol{T2}^{(k+1)}(x)} = \underset{\frac{1}{\boldsymbol{T2}(x)}}{\operatorname{argmin}} \sum_{i>0} \boldsymbol{y}_i^T \left[ \boldsymbol{f}^{(k+1)}(x, TE_i) - \boldsymbol{f}^{(k+1)}(x, TE_0) e^{-\frac{TE_i - TE_0}{\boldsymbol{T2}^{(k)}(x)}} \right]$$

$$+ \sum_{i>0} \frac{\rho}{2} \left| \boldsymbol{f}^{(k+1)}(x, TE_i) - \boldsymbol{f}^{(k+1)}(x, TE_0) e^{-\frac{TE_i - TE_0}{\boldsymbol{T2}^{(k)}(x)}} \right|^2$$

**eq. 8**

$$\boldsymbol{v}^{(k+1)}(x, TE_i) = \underset{\boldsymbol{v}(\mathrm{x}, TE_i)}{\operatorname{argmin}} R[\boldsymbol{v}(x, TE_i), \lambda] + \boldsymbol{z}_i^T [\boldsymbol{v}(x, TE_i) - \boldsymbol{f}^{(k+1)}(x, TE_i)]$$

$$+ \frac{\rho}{2} \left| \boldsymbol{v}(x, TE_i) - \boldsymbol{f}^{(k+1)}(x, TE_i) \right|^2$$

**eq. 9**

$$\boldsymbol{y}_i^{(k+1)} = \boldsymbol{y}_i^{(k)} + \rho [\boldsymbol{f}^{(k+1)}(x, TE_i) - \boldsymbol{f}^{(k+1)}(x, TE_0) e^{-\frac{TE_i - TE_0}{\boldsymbol{T2}^{(k+1)}(x)}}]$$

**eq. 10**

$$\boldsymbol{z}_{\boldsymbol{i}}^{(k+1)} = \boldsymbol{z}_{\boldsymbol{i}}^{(k)} + \rho[\boldsymbol{v}^{(k+1)}(x, TE_i) - \boldsymbol{f}^{(k+1)}(x, TE_i)] \qquad \textbf{eq. 11}$$

The first two subproblems in $\mathbf{eq.}\ \boldsymbol{6}$ and $\mathbf{eq.}\ \boldsymbol{7}$ with respect to the images $\boldsymbol{f}(x, TE_i)$, $i = 0,1, \ldots$, are quadratic optimization problems, and the solutions are computed by solving the linear equation corresponding to the optimality condition.

The subproblem in $\mathbf{eq.}\ \boldsymbol{8}$ with respect to T2 map $\boldsymbol{T2}(x)$ is essentially data fitting to determine the T2 map $\boldsymbol{T2}(x)$ by fitting data $\boldsymbol{f}^{(k+1)}(x, TE_i)$ pixelwise in an exponential decay function form. This was achieved by using a weighted least square (WLS) fitting (eq. 13) of a linear function (eq. 12), where the weight $w_b$ for each term should be inversely proportional to the measurement uncertainty of $\frac{\log\frac{f^{(k+1)}(x,TE_i)}{f^{(k+1)}(x,TE_0)}}{TE_i - TE_0}$. Generally, as this is only a subproblem of the iterative process, it is not necessary to solve this subproblem accurately at each iteration. Hence, a first order approximation of the measurement uncertainty was used as shown in eq. 14, where $\sigma$ is the variance of noise in the measurement of $\boldsymbol{f}^{(k+1)}(x, TE_i)$ and its measurement is described later.

$$\log\left[\boldsymbol{f}^{(k+1)}(x, TE_i)\right] = \log[\boldsymbol{f}^{(k+1)}(x, TE_0)] - \frac{TE_i - TE_0}{\boldsymbol{T2}(x)} \qquad \textbf{eq. 12}$$

$$\frac{1}{\boldsymbol{T2}^{(k+1)}(x)} = \underset{\frac{1}{\boldsymbol{T2}(x)}}{\operatorname{argmin}} \sum_i w_i \left| \log\frac{\boldsymbol{f}^{(k+1)}(x, TE_i)}{\boldsymbol{f}^{(k+1)}(x, TE_0)} + \frac{TE_i - TE_0}{\boldsymbol{T2}(x)} \right|^2 \qquad \textbf{eq. 13}$$

$$w_i \propto \left( -\frac{1}{TE_i - TE_0} \log\left(\boldsymbol{f}^{(k+1)}(x, TE_i) + \sigma\right) + \frac{1}{TE_i - TE_0} \log\left(\boldsymbol{f}^{(k+1)}(x, TE_i) - \sigma\right) \right)^{-1} \qquad \textbf{eq. 14}$$

As for the subproblem defined in **eq. *9***, which is a essentially image domain processing problem, the PnP approach[34] was employed. The PnP method enabled plugging in a powerful image denoising algorithm and its validity has been empirically demonstrated in previous studies in terms of producing high-quality results in various applications[35-37].

For processing of image $\boldsymbol{v}(x, TE_i)$, assuming that the noise in k-space is Gaussian, the noise in the magnitude MR image will still follow a Gaussian distribution after phase correction, or can be approximated as Gaussian distribution when $SNR > 3$[38]. In this study, BM3D method[32] was used in the PnP framework for Gaussian noise removal, and this step is denoted as:

$$\boldsymbol{v}^{(k+1)}(x, TE_i) = BM3D[\boldsymbol{f}^{(k+1)}(x, TE_i), \sigma] \qquad \textbf{eq. 15}$$

where $BM3\text{D}(.)$ stands for the BM3D denoising operation. It needs to be mentioned that the BM3D algorithm requires a hyper-parameter $\sigma$ specifying the Gaussian noise standard deviation. To estimate the $\sigma$ in MR images, which was also used as the measurement uncertainty in weighted linear regression for T2 map fitting, we first removed the background with a threshold-based segmentation algorithm and then estimated the local noise standard deviation of patches with a size of $5 \times 5$ pixels. The mode value of these local noise variance values was computed as an approximation of the noise standard deviation. Then this value was either directly used as parameter $\sigma$ or multiplied with 0.5 as a conservative estimation to preserve more details in the image.

The iterative process in **eq. *6***-eq. 11 continued until convergence, when the mean relative intensity change of $\boldsymbol{f}(x, TE_i)$ in two successive iteration steps is less than a threshold $\epsilon$. Note that due to the modifications to the ADMM algorithm and the application of PnP framework, theoretical convergence of this iterative process was not guaranteed. $\epsilon$ was set as $1\%$ in this study.

The algorithm used to solve the joint reconstruction and data fitting problem is summarized in **Algorithm 1.** Solving the problem in eq. *3*. The algorithm was implemented in Python 3.8, and computation was performed on a workstation with an Intel(R) Xeon(R) Gold 6230R CPU of 2.1GHz frequency. The other adjustable parameter in this algorithm, $\rho$, was quite robust and was set as 0.5 for all the applications.

---

**Algorithm 1.** Solving the problem in eq. 3.

---

**Input:** K-space data $\boldsymbol{g}(z, TE_i)$ with at least 3 different $TE$ values, sampling matrix $\boldsymbol{S}$ and phase matrix $\boldsymbol{A}$

**Parameters:** $\rho, \epsilon$

**Initialize:**

$$\boldsymbol{v}^{(0)}(x, TE_i) = \boldsymbol{f}^{(0)}(x, TE_i) = \boldsymbol{A}^{-1}\boldsymbol{F}^{-1}\boldsymbol{S}\boldsymbol{g}(z, TE_i),$$

$$\frac{1}{\boldsymbol{T2}^{(0)}(x)} = \underset{\frac{1}{\boldsymbol{T2}(x)}}{\operatorname{argmin}} \sum_i w_i \left| \log \frac{\boldsymbol{f}^{(0)}(x, TE_i)}{\boldsymbol{f}^{(0)}(x, TE_0)} + \frac{TE_i - TE_0}{\boldsymbol{T2}(x)} \right|^2,$$

$$\boldsymbol{y}_i^{(0)} = \boldsymbol{z}_i^{(0)} = 0, \text{ and } k = 0$$

**While:**

$$\text{mean}\left(\frac{\left|\boldsymbol{f}^{(k+1)}(x, TE_i) - \boldsymbol{f}^{(k)}(x, TE_i)\right|}{\boldsymbol{f}^{(k)}(x, TE_i)}\right) > \epsilon \text{ or } k = 0$$

**Do:**

$k = k + 1$

Solve problems in eq. 6 and eq. 7 using CGLS algorithm

Update $\boldsymbol{T2}$ by pixelwise data fitting eq. ***13***

Update $\boldsymbol{v}$ using BM3D algorithm in eq. 15

Update $\boldsymbol{y}_i$, $\boldsymbol{z}_i$ by eq. 10 and eq. 10

**End while**

**Return:** $\boldsymbol{f}^* = \boldsymbol{f}^{(k+1)}$ and $\boldsymbol{D}^* = \boldsymbol{D}^{(k+1)}$

---

### 2.C. Evaluation

The joint reconstruction algorithm was evaluated in phantom and patient studies. The study protocol was approved by our Institutional Review Board and written informed consent was waived from each subject. T2W imaging data sets were acquired on a phantom (Essential System Phantom for Relaxometry, Model 106, CaliberMRI, Boulder, CO) consisting of 14 vials with different concentration of $MnCl_2$ solution providing a T2 range from 8 to 850 $ms$. measured on 1.5T at 20°C. MR images were acquired on a 1.5T clinical MRI scanner (Ingenia Ambition X, Philips Healthcare, Netherlands) with a 20-channel brain coil. In phantom study, T2W imaging was acquired using a multi-echo FSE sequence following the vendor-provided protocol: field-of-view (FOV) $= 250 \times 250\, mm^2$, slice thickness/gap = 6/0 mm, matrix size = $256 \times 256$, repetition time (TR) = 5000 ms, eleven TEs from 11 to 176 ms with 11 ms step, bandwidth = 170 Hz/pixel, number of excitation (NEX) = 1, echo train length = 16, acquisition time = 21 min 25 $sec$. In subject scan, T2W imaging was acquired using a GraSE sequence with the following parameters: field-of-view (FOV) $= 256 \times 207\, mm^2$, slice thickness/gap = 3/9 mm, matrix size = $400 \times 400$, repetition time (TR) = 1892 ms, 32 TEs from 14.4 to 461.6 ms with 14.4 ms step, bandwidth = 691 Hz/pixel, number of excitation (NEX) = 2, echo train length = 32, echo planar factor = 5, acquisition time =2 min 42 sec.

For both phantom and subject scans, each T2W image was converted to the corresponding k-space data through FFT. Furthermore, we tested the effectiveness of the proposed joint k-TE reconstruction method in reconstructing under-sampled k-space data, in which the central 10% k-space data was kept while 25% and 33% of the rest 90% k-space data was randomly discarded. A CS-based reconstruction algorithm[39] was used as the conventional reconstruction method to reconstruct the T2W image from the under-sampled data. Each MR image was reconstructed under the regularization of 1D Total Variation (TV) as shown in eq.16, where $\boldsymbol{F}_{phase}^{1D}, \boldsymbol{F}_{freq}^{1D}$ and $TV_{phase}^{1D}$ denoted 1D Fourier transform applied along phase direction, 1D Fourier transform applied along frequency direction, and 1D TV applied along phase encoding direction, respectively.

$$\boldsymbol{f}^* = \underset{\boldsymbol{f}}{\operatorname{argmin}} \sum_i |\boldsymbol{S}\boldsymbol{F}_{freq}^{1D}\boldsymbol{F}_{phase}^{1D}\boldsymbol{A}\boldsymbol{f}(x, TE_i) - \boldsymbol{g}(z, TE_i)| + TV_{phase}^{1D}[\boldsymbol{F}_{phase}^{1D}\boldsymbol{A}\boldsymbol{f}(x, TE_i)] \quad \textbf{eq. 16}$$

To numerically evaluate the performance of the proposed joint k- TE reconstruction method on both fully sampled and under-sampled phantom data, mean and standard deviation of T2 measurements were calculated in each vial of the phantom with ground truth T2 values, as well as in three regions-of-interests (ROIs) including white matter (WM), gray matter (GM), and ventricle on subject images. The joint reconstruction results were compared with those computed using the conventional 2-step WLS fitting method on the fully sampled data.

## 3. RESULTS

In all of the fully sampled and under-sampled datasets of phantom and subject scans, the proposed algorithm converged within 10 iterations in fully sampled data, as shown in Figure. 1. It took slightly more iterations for the algorithm to converge on the under-sampled data due to the ill-defined problem. The mean relative change (MRC) of the image intensity increased at the beginning of iterations and we started to check it from the $4^{th}$ iteration.

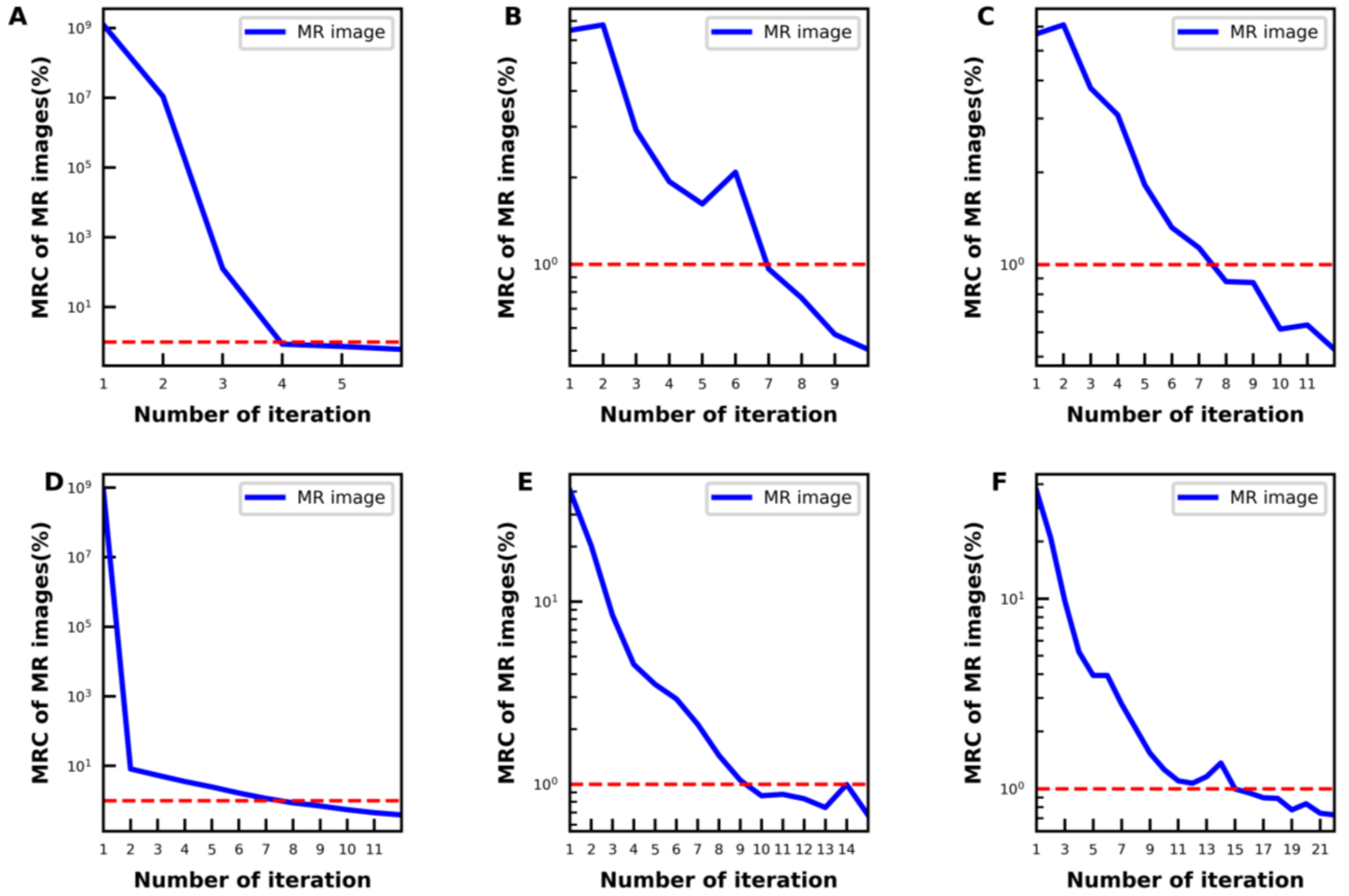

**Figure. 1.** The convergence plot of the proposed algorithm, shown as the mean relative change (MRC) along with the number of iterations in fully sampled phantom data (**A**), 25% under-sampled phantom data (**B**), 33% under-sampled phantom data (**C**), fully sampled patient data (**D**), 25% under-sampled patient data (**E**), and 33% under-sampled patient data (**F**). The horizontal dashed red line in each subplot shows the stopping criteria in each data set.

### 3.A. Image Quality Evaluation

Image quality comparison between fully sampled, under-sampled joint reconstruction methods, and full sampled conventional method were shown in **Fig. 2** (phantom) and **Fig. 3** (subject). The joint reconstruction method demonstrated improved image quality with less noise on T2W images at low, medium, and high TEs and the corresponding T2 map in both fully sampled and under-sampled data sets. The proposed algorithm outperformed the conventional CS method in under-sampled image reconstruction by removing the aliasing artifacts more effectively.

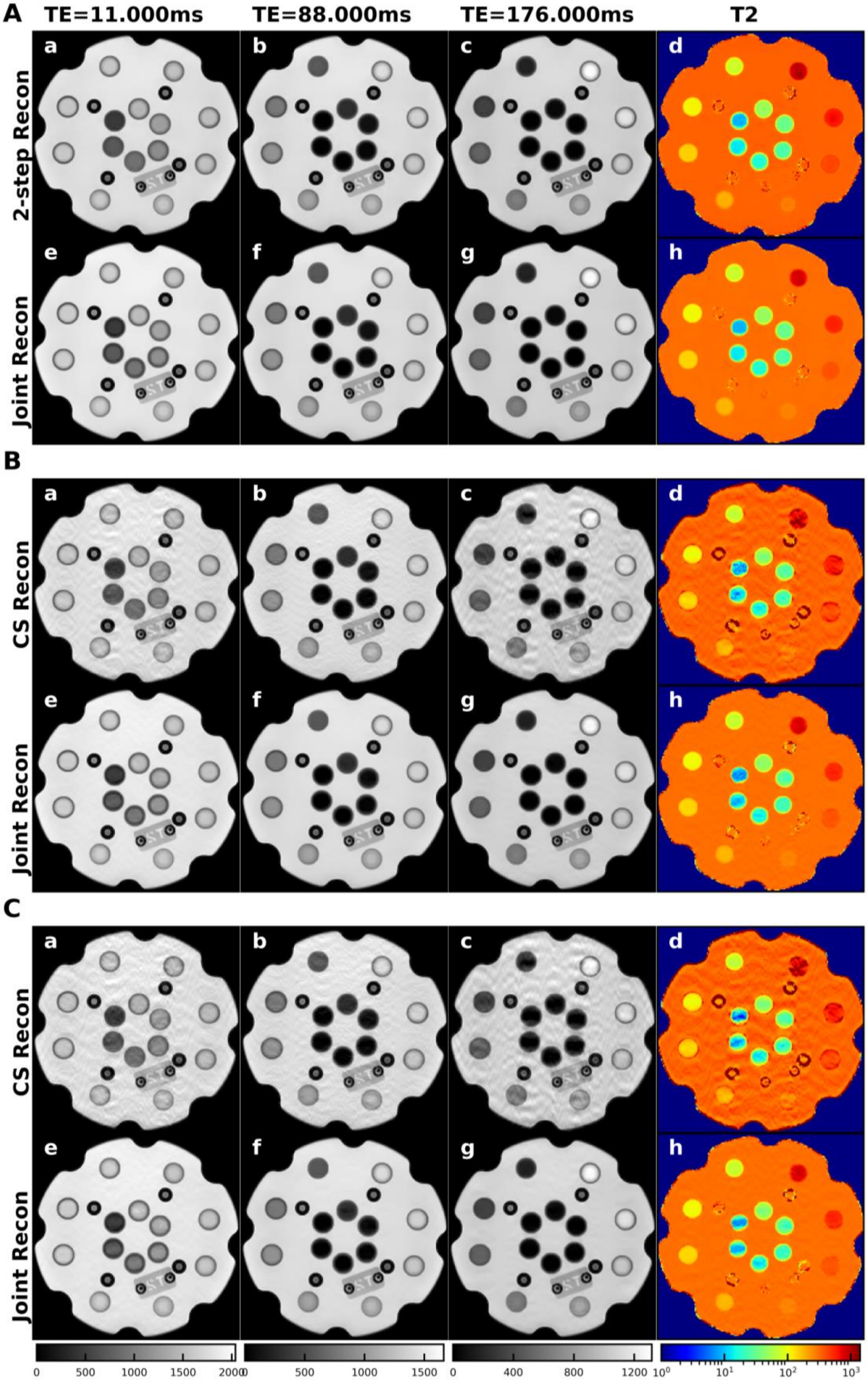

**Figure 2**. The reconstructed phantom images by the conventional 2-step and proposed joint reconstruction method in the fully sampled (A), 25% under-sampled (B) and 33% under-sampled (C) phantom data set. In all data sets, T2W MR images and the corresponding T2 map reconstructed by the k-TE joint reconstruction (Joint Recon) method improved image quality compared with those reconstructed by the conventional 2-step (2-step Recon) method for fully sample data and by the compressed-sensing (CS Recon) based method for under-sampled data.

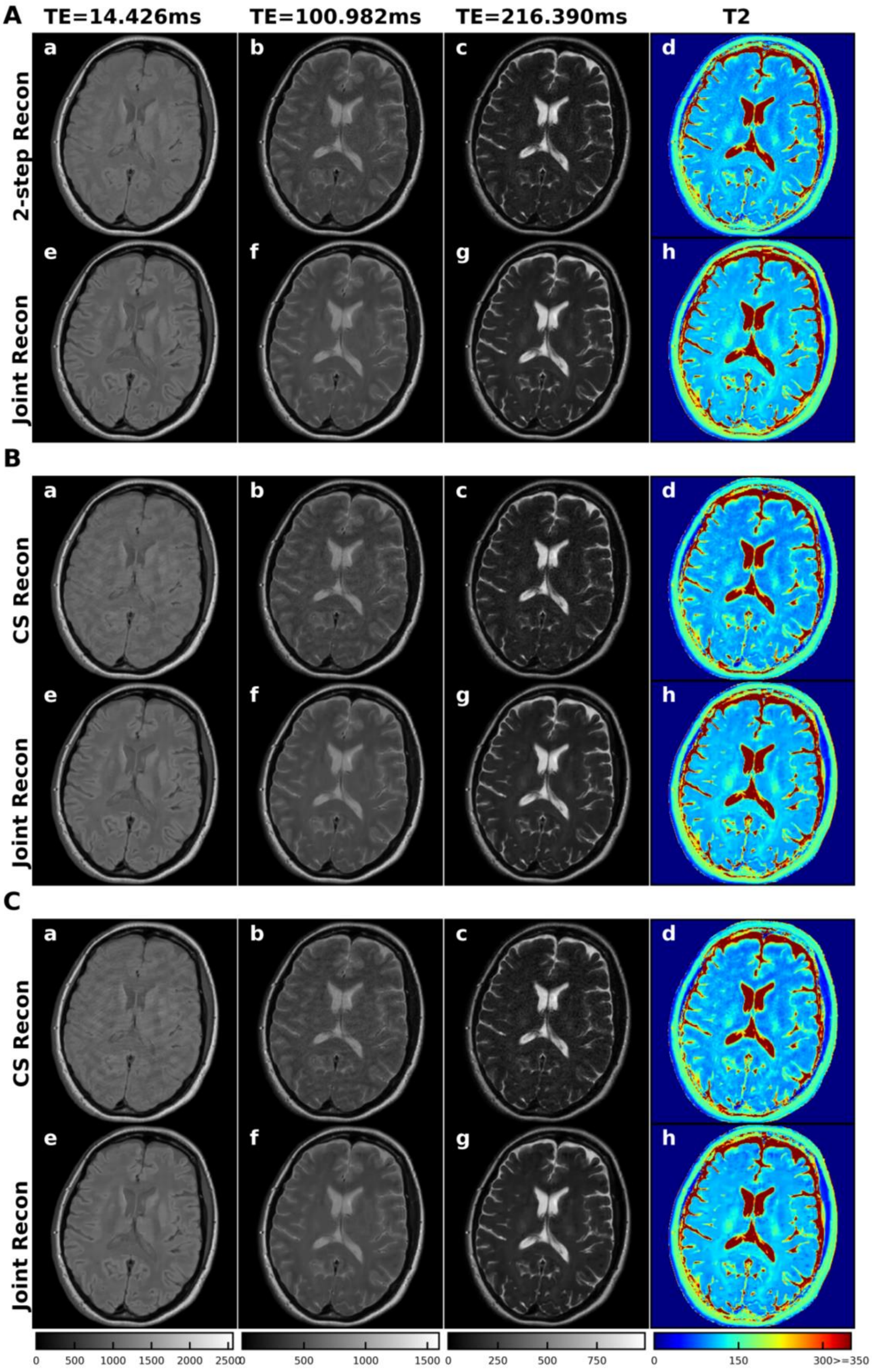

**Figure 3.** The reconstructed brain images by the conventional and proposed method in the fully sampled (A), 25% under-sampled (B) and 33% under-sampled (C) subject brain data. In all data sets, T2W MR images and the corresponding T2 map reconstructed by the k-TE joint reconstruction (Joint Recon) method improved image quality compared with those reconstructed by the conventional 2-step (2-step Recon) method for fully sampled data and by the compressed-sensing (CS Recon) based method for under-sampled data.

### 3.B. Quantitative T2 Measurement Evaluation

The accuracy of T2 measurements in the Relaxometry phantom by the conventional and proposed algorithms on both fully sampled and under-sampled data was shown in **Table. 1**. The vendor provided T2 values for each vial in the phantom were used as the ground truth. The mean and standard deviation of the pixel-wise T2 values were calculated within a ROI manually placed in each vial of the phantom, theoretically all pixels within each ROI should have the same T2 value.

In the phantom study, the joint reconstruction method applied to both fully sampled and under-sampled data provided comparable mean T2 measurements as the conventional method in fully sampled dataset, and they were consistent with the gold standard T2 values. In addition, the variation in pixel-by-pixel T2 measurements was greatly reduced (i.e., less standard deviations) using the joint reconstruction method in the full sample data set comparison. The mean T2 values of the under-sampled reconstruction by the proposed algorithm showed minimal differences compared with those reconstructed from the fully sampled data, however, the variations in T2 measurements of under-sampled data were higher than those of fully sampled reconstruction.

**Table 1**. Comparison of mean and standard deviation of T2 values reconstructed by the conventional 2-step reconstruction on fully sampled phantom data (Full 2-step Recon) and the proposed joint reconstruction method on fully sampled phantom data (Full Joint Recon), 25% under-sampled phantom data (25% Joint Recon) and 33% under-sampled phantom data (33% Joint Recon).

| Ground Truth ($ms$) | | 8.75 | 12.8 | 17.9 | 26.1 | 34.3 | 53.0 | 82.2 |
|---|---|---|---|---|---|---|---|---|
| Full 2-step Recon | mean | 8.44 | 12.4 | 16.1 | 24.0 | 31.7 | 48.6 | 75.7 |
| | std | 1.67 | 1.23 | 0.924 | 1.01 | 0.775 | 0.916 | 0.862 |
| Full Joint Recon | mean | 8.40 | 12.4 | 16.0 | 23.8 | 31.2 | 47.6 | 73.8 |
| | std | 1.20 | 0.918 | 0.756 | 0.798 | 0.578 | 0.945 | 0.911 |
| 25% Joint Recon | mean | 7.45 | 10.6 | 14.5 | 21.3 | 26.6 | 46.4 | 73.1 |
| | std | 2.55 | 3.45 | 2.24 | 1.77 | 4.35 | 3.22 | 1.06 |
| 33% Joint Recon | mean | 11.7 | 13.5 | 15.5 | 20.4 | 26.3 | 44.5 | 72.8 |
| | std | 7.98 | 9.53 | 6.86 | 3.44 | 6.29 | 6.78 | 1.73 |

*The unit is $ms$ for all mean and std of ROI T2

**Table 2 continued**.

| Ground Truth ($ms$) | | 116 | 167 | 194 | 323 | 479 | 692 | 853 |
|---|---|---|---|---|---|---|---|---|
| Full 2-step Recon | mean | 109 | 147 | 187 | 278 | 421 | 641 | 954 |
| | std | 1.53 | 2.03 | 3.28 | 6.38 | 13.1 | 29.8 | 85.3 |
| Full Joint Recon | mean | 105 | 141 | 179 | 263 | 385 | 556 | 797 |
| | std | 1.34 | 1.18 | 2.33 | 4.60 | 5.31 | 15.1 | 37.1 |
| 25% Joint Recon | mean | 104 | 140 | 179 | 264 | 376 | 539 | 781 |
| | std | 1.33 | 2.43 | 4.15 | 8.23 | 11.5 | 28.9 | 69.3 |
| 33% Joint Recon | mean | 104 | 139 | 179 | 262 | 376 | 535 | 786 |
| | std | 1.46 | 2.54 | 4.57 | 9.37 | 12.3 | 35.7 | 79.7 |

In human brain data set, the mean and standard deviation of ADC values were calculated within three selected ROIs (white matter, gray matter, and CSF: cerebrospinal fluid) as shown in **Table. 2**. Since the noise distribution in MR images can be approximated as Gaussian distribution, it had little impact on the mean ADC value reconstructed by the conventional 2-step algorithm and thus we used the conventional reconstructed ADCs as ground truth. In all three ROIs, the joint reconstructed ADC values were comparable to the ground truth while showing higher consistency (i.e., less standard deviation).

**Table 2.** Comparison of mean and standard deviation of T2 values reconstructed by the conventional 2-step reconstruction on fully sampled (Full 2-step Recon) and the proposed joint reconstruction method on fully sampled (Full Joint Recon), 25% under-sampled (25% Joint Recon), and 33% under-sampled data (33% Joint Recon) of a brain scan.

| | | CSF | White Matter | Gray Matter |
|---|---|---|---|---|
| Full 2-step Recon | mean | 1716 | 94.96 | 117.7 |
| | std | 177.6 | 4.266 | 9.220 |
| Full Joint Recon | mean | 1733 | 104.1 | 124.8 |
| | std | 139.5 | 2.305 | 7.974 |
| 25% Joint Recon | mean | 1729 | 102.1 | 124.8 |
| | std | 146.4 | 2.156 | 7.156 |
| 33% Joint Recon | mean | 1731 | 101.6 | 128.3 |
| | std | 140.4 | 1.550 | 8.367 |

*The unit is $ms$ for all mean and std of ROI T2

## 4. DISCUSSION

In this work, we developed a novel optimization method for joint reconstruction of T2W MR images and T2 map by exploiting constraints simultaneously from k-space and TE-space that used regularizations in image domain and self-consistency condition of the T2 weighted exponential signal decay form. This proposed algorithm improved image quality in T2W images and T2 maps, and reduced variations in pixel-by-pixel T2 measurements.

Technical highlights of the proposed algorithm reside in the synergy of k-TE space constraints in MR image reconstruction and T2 fitting through multiple iterations. Considering the mono-exponential decay relationship between T2W signals and T2 value, we incorporated a constraint in TE space to the optimization problem, which enforced the mono-exponential decay calibration through MR signals acquired at different TEs with the signal decay coefficient being $\frac{1}{T2}$. This TE-space constraint was based on that all k-space data acquired at different TEs rather than single k-space data acquired at single TE. The detailed updating formulas through each iteration to reconstruct a T2W MR image ($i = 0$ and $i > 0$) were shown in **Eq. *17*** and **Eq. *18***.

$$\boldsymbol{f}_0 = \frac{\frac{\boldsymbol{g}^T\boldsymbol{S}\boldsymbol{F}^*\boldsymbol{A}^* + \boldsymbol{g}^H\boldsymbol{S}\boldsymbol{F}\boldsymbol{A}}{2} + \sum_{i>0}\left(\boldsymbol{y}_i^T e^{-\frac{TE_i - TE_0}{\boldsymbol{T2}}} + \rho \boldsymbol{f}_b^T e^{-\frac{TE_i - TE_0}{\boldsymbol{T2}}}\right) + \boldsymbol{z}_0^T + \rho \boldsymbol{v}_0^T}{\frac{\boldsymbol{A}^H\boldsymbol{F}^*\boldsymbol{S}\boldsymbol{F}\boldsymbol{A} + \boldsymbol{A}\boldsymbol{F}\boldsymbol{S}\boldsymbol{F}^*\boldsymbol{A}^*}{2} + \rho + \rho * \sum_{I>0} e^{-2\frac{TE_i - TE_0}{\boldsymbol{T2}}}} \qquad \textbf{Eq. 17}$$

$$\boldsymbol{f}_{i,i>0} = \frac{\frac{\boldsymbol{g}^T\boldsymbol{S}\boldsymbol{F}^*\boldsymbol{A}^* + \boldsymbol{g}^H\boldsymbol{S}\boldsymbol{F}\boldsymbol{A}}{2} - \boldsymbol{y}_i^T + \rho \boldsymbol{f}_0^T e^{-\frac{TE_i - TE_0}{\boldsymbol{T2}}} + \boldsymbol{z}_i^T + \rho \boldsymbol{v}_i^T}{\frac{\boldsymbol{A}^H\boldsymbol{F}^*\boldsymbol{S}\boldsymbol{F}\boldsymbol{A} + \boldsymbol{A}^T\boldsymbol{F}\boldsymbol{S}\boldsymbol{F}^*\boldsymbol{A}^*}{2} + \rho + \rho} \qquad \textbf{Eq. 18}$$

The numerator contained terms for data fidelity (i.e., k-space constraint), TE-space constraint, and the smoothness constraint, aside from the Lagrange multiplier terms about $\boldsymbol{z}$ and $\boldsymbol{y}$. The denominator was the normalization term. The meanings of these equations are interpretable. For MR image at $TE_0$, each MR image with $TE_i, i > 0$ from the last iteration estimated an expected MR image at $TE_0$ based on the exponential decay relationship. Then these expected images were averaged together with the image generated based on data fidelity, the image after applying smoothing constraints and those from multipliers. Similarly, for updating an MR image at $TE_i, i > 0$, the expected MR image given by $\boldsymbol{f}_0$ was involved. Overall, as the iterations continued, each T2W MR image was estimated by calibrating data from all k-space data at different TEs rather than relying on its only k-space data. The iterative process effectively removed noise that was randomly distributed in each MR image, and thus improve the image quality and the accuracy of T2 measurements.

In addition to the k-TE space constraints, the smoothness constraint was also considered in MR images. BM3D was incorporated in our algorithm using a PnP method as the smoothness regularization for MR images to remove noise with Gaussian distribution. We also tested Rician-based BM3D method instead assuming the noise in MR images follows Rician distribution but found out both BM3D and Rician-based BM3D gave similar results. This could be because the Rician-distributed noise can be approximated as Gaussian noise in most of our data sets. We also tested apply the BM3D spatial regularizations in T2 map, hoping to further improve the jointly reconstructed image quality. However, this approach tended to over-smooth the reconstructed images, which may be due to the unknown noise distribution in T2 map.

It is not able to mention that the joint reconstruction algorithm was more effective in removing the aliasing artifacts on the under-sampled data compared with CS-based algorithm, mainly because of the combined application of TE-space constraints, rather than the application of the denoising tool BM3D, but. To validate the effectiveness of combining both k-space and TE-space constraints in the proposed algorithm, we tested only applying the k-space constraint regularized with BM3D solely to reconstruct the fully sampled and under-sampled phantom data by minimizing eq. 19:

$$\{\boldsymbol{f}^*\} = \underset{\boldsymbol{f}}{\operatorname{argmin}} \sum_i |\boldsymbol{SFA}(x, TE_i)\boldsymbol{f}(x, TE_i) - \boldsymbol{g}(z, TE_i)|^2 + R[\boldsymbol{f}(x, TE_i), \lambda] \qquad \textbf{eq. 19}$$

The optimization process still exploited the ADMM and PnP algorithms, which was similar to that of the proposed joint k-TE reconstruction algorithm. In this test, MRC failed to converge in all phantom data sets and thus the iteration was manually stopped at the 4$^{th}$ iteration. As shown in **Fig. 4**. the approach cannot remove noise in the fully sampled data set, especially for T2W MR images at high TE. The results became worse in under-sampled data set, where the aliasing patterns with high spatial correlation were more obvious. This test demonstrated the effectiveness and necessity of joint application of the constraints from both k-space and TE-space.

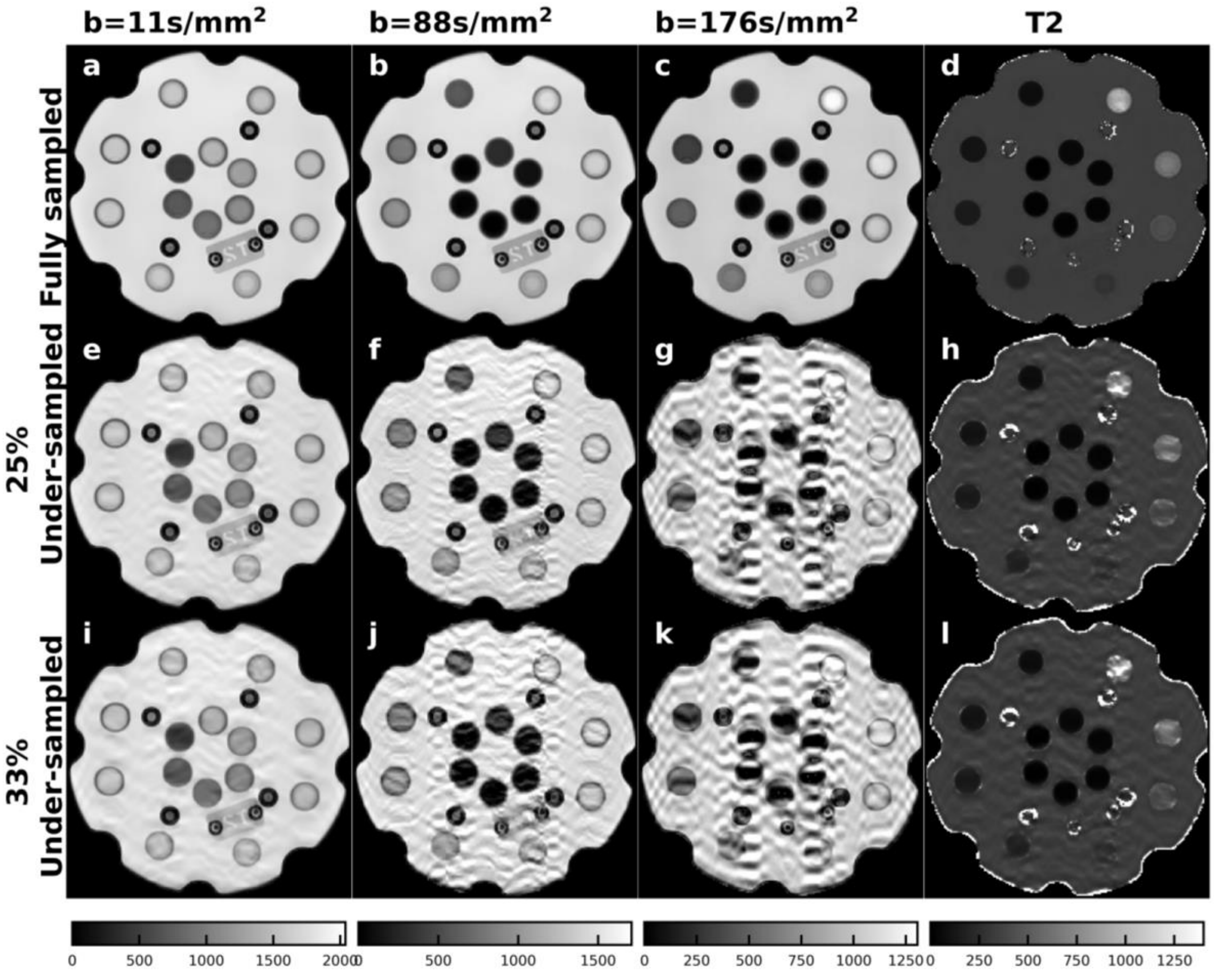

**Figure 4.** Images reconstruction by only applying BM3D regularized k-space constraint in the fully sampled (a-f), 25% under-sampled (e-f) and 33% under-sampled (i-l) phantom data.

The adjustable parameters in the joint reconstruction method included the model parameter $\rho$ in **eq. *5*** brought by augmented Lagrange method and the threshold $\epsilon$ for convergence judgement. Among them, the model parameter $\rho$ did not impact the final convergence result but only affected the rate and stability of convergence. It can be set to a larger value when the convergence process was not stable at the expense of having more iterations before reaching convergence. The threshold $\epsilon$ was used to stop the algorithm at a proper iteration and needed to be fine-tuned based on different data sets. In this study, the threshold $\epsilon$ was set as $1\%$ for both phantom and subject data sets.

There are several limitations in our work. First, the threshold of MRC for stopping the iterations may vary in different data sets. Currently a conservative way to find the best threshold was to first run the algorithm with a small threshold and then select a proper threshold that preserves as much fine structures as possible while having the noise suppressed to overcome the over smoothing issue. Second, the reconstruction time for the proposed algorithm with weighted least square fitting is long (5 mins each iteration for each slice with $400 \times 400$ pixels and 32 TEs) compared with the conventional FFT method. Over half the time was spent on noise variance estimation, which can be accelerated using high performance GPU computing.

Future work will include extending the joint k-TE reconstruction method to more complex models such as 3-parameter T2 model fitting model that takes into account the effect of imaging pluses[40]. Also, a more robust algorithm will be implemented to estimate the parameter $\sigma$ for BM3D in order to avoid fine tuning the threshold $\epsilon$. Taking one step further, BM3D might not be the optimal choice to be plugged in as the denoising regularization tool, more flexible tools, such as a neural network, can be integrated to achieve better denoising effect and image reconstruction.

## 5. CONCLUSION

We developed a novel joint k-TE space optimization algorithm for simultaneous T2W MR image and T2 map reconstruction. In this method, the k-space constraint enforced data fidelity, and TE-space constraint enabled information to be shared among MR images at different TEs and the corresponding T2 map. Our algorithm improved the image quality of T2W MR images and T2 maps with better SNR, increased the stability of pixelwise T2 measurements compared with the conventional magnitude image-based 2-step signal fitting method.